\documentclass[preprint,showpacs,preprintnumbers,amsmath,amssymb]{revtex4}
\usepackage{graphicx}
\usepackage{dcolumn}
\usepackage{bm}
\begin{document}

\title{
Observation of photoinduced phase transition in phase-separated Pr$_{0.55}$(Ca$_{1-y}$Sr$_{y}$)$_{0.45}$MnO$_3$
thin films via x-ray photoemission spectroscopy
}

\author{K. Takubo}
\affiliation{Department of Physics and Department of Complexity Science and Engineering,
University of Tokyo, 5-1-5 Kashiwanoha, Chiba 277-8581, Japan}
\author{J.-Y. Son}
\affiliation{Department of Physics and Department of Complexity Science and Engineering,
University of Tokyo, 5-1-5 Kashiwanoha, Chiba 277-8581, Japan}%
\author{T. Mizokawa}
\affiliation{Department of Physics and Department of Complexity Science and Engineering,
University of Tokyo, 5-1-5 Kashiwanoha, Chiba 277-8581, Japan}
\author{N. Takubo}
\affiliation{
Research Center for Advanced Science and Technology {\it (RCAST)}, University of Tokyo, Tokyo 153-8904, Japan}
\author{K. Miyano}
\affiliation{
Research Center for Advanced Science and Technology {\it (RCAST)}, University of Tokyo, Tokyo 153-8904, Japan}

\date{\today}

\begin{abstract}

Perovskite manganite thin films,
Pr$_{0.55}$(Ca$_{1-y}$Sr$_y$)$_{0.45}$MnO$_3$, have been studied using x-ray photoemission
spectroscopy in order to clarify the consequence of the competition 
between ferromagnetic metal (FM) and charge-orbital ordered insulator (COOI). 
Films with $y$ = 0.40 undergo uniform paramagnetic insulator to FM
transition. 
On the other hand, in films with $y$ = 0.25, the composition near the bicritical point, 
phase separation of COOI and FM domains is indicated by the spectral change
below 125 K. Interestingly, between 50 K and 70 K, the visible laser 
illumination transfers the COOI-like spectra obtained in cooling process 
to the FM-like spectra obtained in warming process. This indicates
that the photoinduced IMT is governed by the increase of the FM volume fraction
and is deeply related to the phase separation between the FM and COOI states.

\end{abstract}

\maketitle

Photoinduced phase transition (PIPT) has recently attracted considerable 
attention. A wide variety of PIPT has been reported such as 
the spin transition in iron complexes \cite{Decurtins},
magnetization in cobalt-iron cyanides \cite{Sato},
valence transition in Cs$_2$Au$_2$Br$_6$ \cite{Liu},
order to disorder phase transition in InSb \cite{Lindenberg,Sokolowski},
insulator to metal transition (IMT) in (EDO-TTF)$_2$PF$_6$ \cite{Chollet} 
and VO$_2$ \cite{Cavalleri}, and so on. 
Among them, photoinduced IMT in perovskite manganites is unique in
that despite the orders of magnitude resistance drop as a result of
photo-irradiation, it has been indicated that the resulting conduction
path is only filamentary \cite{Miyano}. 
This is in line with the proposal of the colossal magnetoresistance
(CMR) at low temperatures being percolative phenomena in a
metal-insulator mixed state; a relatively small magnetic field
can cause enormous resistance drop simply because the magnetic field
needs to destroy only small fraction of intervening insulating phase between metallic islands. This is the scenario called CMR1 \cite{aliaga}.

Although the phenomenon is very clear superficially, the role the
photoexcitation plays in the IMT is not clear at all. Let us recall
that the `evidences' of the photoinduced IMT in manganites
so far are the macroscopic conductivity jump \cite{Miyano,NTakubo} and the increase
in the magnetization \cite{okimoto}, none of which proves that the new photoinduced
phase is indeed metallic. It should be pointed out that simple photoinduced
redistribution or coalescence of metallic islands can establish
connectivity of metallic phase in a two-phase coexisting state. A
ferromagnetic state can well be insulating in manganites \cite{endoh}.

In order to clarify these questions, we performed x-ray photoemission
spectroscopy (XPS) measurements on manganite thin films under
photo-irradiation. By observing the electronic density of states, we found 
the first direct evidence of the 
photoinduced metallic phase. The PIPT, however,
occurs only within a hysteresis loop with a clear temperature
cutoff. This indicates that the presence of competing orders
separated by a first-order phase transition in the
vicinity of a multicritical point is essential for the PIPT. The two
distinct ordered states need not be coexisting before
photo-irradiation, unlike the CMR1 scenario, but the parameter space
for the true PIPT is rather small, which conforms to our experience
that the photoinduced IMT is limited to a small class of manganites.

Thin films of Pr$_{0.55}$(Ca$_{1-y}$Sr$_y$)$_{0.45}$MnO$_3$ (PCSMO) coherently
grown on (LaAlO$_3$)$_{0.3}$(SrAl$_{0.5}$Ta$_{0.5}$O$_3$)$_{0.7}$ (LSAT) (011) substrate
have been employed as an archetypal example \cite{NTakubo}.
The thickness is about 80 nm.
Two compositions ($y$ = 0.25 and 0.40) were chosen. The $y$ = 0.40 sample
shows simple paramagnetic insulator (PI) to ferromagnetic metal
(FM) transition on cooling while the $y$ = 0.25 sample is
located near a bicritical point and undergoes PI to charge-orbital
ordered insulator (COOI) transtition followd by a transition to FM.
The photoemission spectroscopy has been performed using a JPS9200 
spectrometer equipped with a monochromatized Al${K\alpha}$
x-ray source ($h\nu$ = 1486.6 eV). The total resolution was $\sim$0.6 eV.
The pressure of the spectrometer was $\sim$1$\times$10$^{-7}$ Pa 
during the measurement.
The thermal cycles of XPS measurements were performed repeatedly to
confirm reproducibility. 
A Nd:YAG pulsed laser provided optical excitation of 2.3 eV
(532 nm) at a repetition rate of 30 Hz with the pulse width of 10 ns.
The energy per pulse was 1.0 mJ and the beam diameter was about 
4 mm corresponding to the density of $1.7\times 10^{16}$cm$^{-2}$\ photons per pulse.

\begin{figure}
\includegraphics[width=8cm,clip]{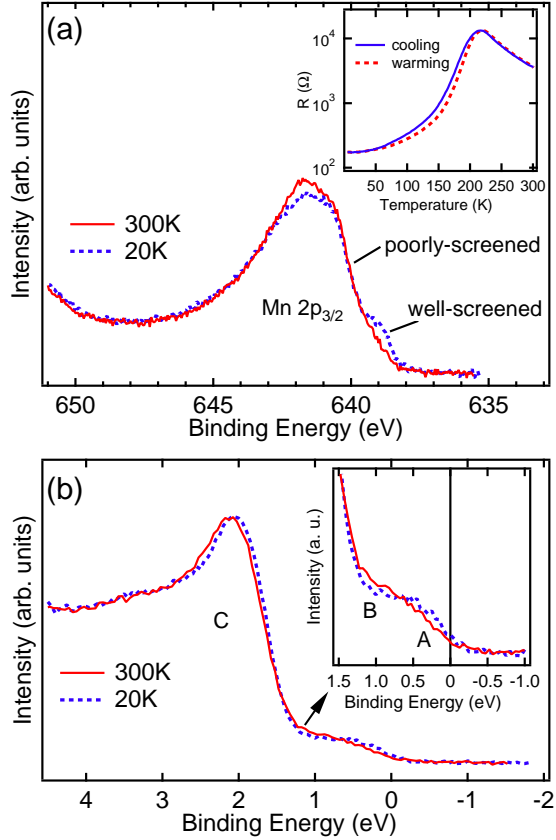}%
\caption{
(Color online). (a) Mn $2p_{3/2}$ and (b) valence-band XPS data of
PCSMO ($y$ = 0.40) thin film 
at 35 K and 300 K. The inset of the upper panel shows the resistance of the film.
}
\label{y40}
\end{figure}

Figure \ref{y40} shows the (a) Mn $2p_{3/2}$ and
(b) valence-band spectra of $y$ = 0.40 film.
The spectra are normalized to the integrated intensities from 636.0 eV
to 648.0 eV for the Mn $2p_{3/2}$ 
and from $-$1.0 eV to 4.0 eV for the valence-band, respectively.
The PI to FM transition occurs at $\sim$200 K [inset of Fig. \ref{y40} (a)]. 
The XPS data show dramatic change in going from PI to FM.
While the Mn $2p_{3/2}$ spectrum of the PI ($\sim$300 K) has only the poorly-screened feature ($\sim$641 eV), 
the spectrum of the FM ($\sim$20 K)  additionally has the well-screened feature 
at the lower binding energy side ($\sim$639 eV) due to the Mn $3d$ carriers.
The well-screened feature reflects the core-hole screening through 
 the  density of states at the Fermi level [$D(E_F)$]\cite{Horiba, Tanaka}.
This is in accordance with the weight enhancement of $D(E_F)$
(structure A) through the weight transfer from  structure B ($\sim$1.0
eV) in the valence
band spectra with decreasing temperature (Fig. \ref{y40} (b)).
Structures A and B are the coherent and incoherent bands of the Mn $3d$ $e_g$ states,
while the Mn $3d$ $t_{2g}$ and Pr $4f$ states sit around $\sim$ 2 eV (structure C)\cite{Chainani,Sekiyama}.
The obvious enhancement of $D(E_F)$ in the metallic phase
suggests three-dimensional metallic character.
 
\begin{figure}
\includegraphics[width=8cm,clip]{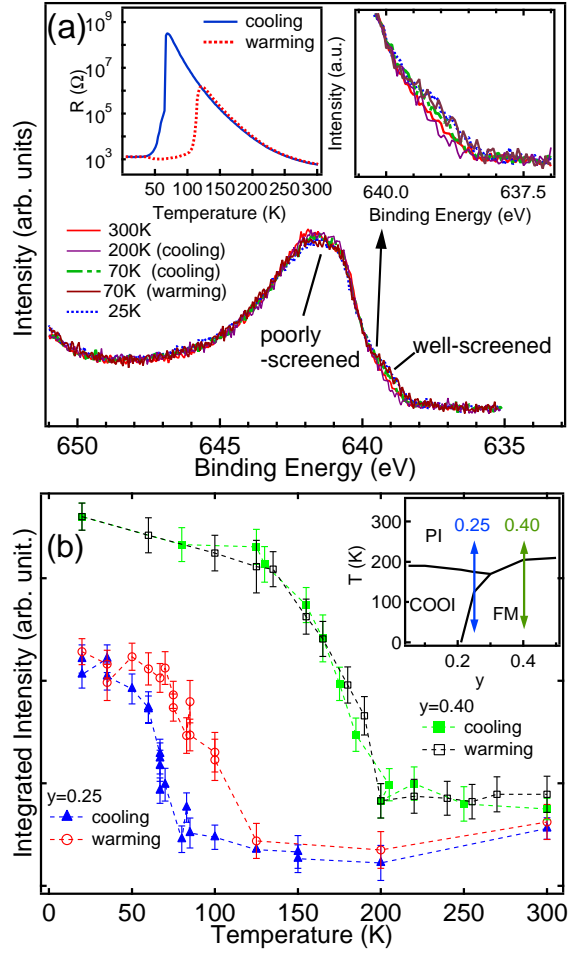}%
\caption{(Color online). (a) Mn $2p_{3/2}$ XPS data of
PCSMO ($y$ = 0.25) thin film at various temperatures. The left inset shows the resistance of the film.
In the right inset, the spectra of the well-screened feature are expanded.
(b) Temperature dependence of the integrated intensity from 638.0 eV to 640.0 eV for the PCSMO XPS data.
The (blue) closed triangles and (red) open circles indicate the cooling and warming process for $y$ = 0.25, respectively.
The (green) closed squares show the plots for $y$ = 0.40. The inset of lower panel shows the phase diagram of PCSMO.    
}
\label{y25Mn}
\end{figure}

Figure \ref{y25Mn} (a) gives the Mn $2p_{3/2}$ XPS data of $y$ = 0.25 film.
The spectra are normalized in the same way as for $y$ = 0.40. 
Although the well-screened structure of Mn $2p_{3/2}$ ($\sim$639 eV) 
for $y$ = 0.25 also appears at low temperature, 
it is weaker compared to $y$ = 0.40.
For the quantitative evaluation,
the integrated intensity from 638.0 eV to 640.0 eV, namely, the area of the well-screened feature, 
is plotted against temperature in Fig. \ref{y25Mn} (b) for both $y$ =
0.40 and 0.25. 
For $y$ = 0.40, in the cooling process, the integrated intensity 
(green closed squares) starts to increase at the IMT temperature
($\sim$ 200 K) of the resistance.
The onset of integrated intensity matches 
the resistance drop, indicating that the PI to FM transition 
takes place uniformly and no phase separation appears in the film.

However, for $y$ = 0.25 there is a discrepancy between the temperature
dependence of the integrated intensity and that of the resistance. In
the cooling process, 
although the integrated intensity gradually increases below 125 K (blue closed triangles),
the resistance keeps increasing down to $\sim$ 67 K.
This indicates that the FM domains start to appear in the COOI phase below 125 K 
in the cooling process.
On the other hand, in the warming process (red open circles) the integrated intensity
gradually decreases above 50 K but the resistance stays at the low metallic value up to 110 K.
This suggests that the COOI domains start to grow in the metallic phase. 
Moreover there is a distinct thermal hysteresis between the cooling
and warming curves of the integrated intensity.
The hysteretic temperature region is thus also characterized by the
two-phase coexistence of FM and COOI, similar to the previous report
about La$_{1-x-y}$Pr$_y$Ca$_x$MnO$_3$ in Ref. \onlinecite{Sarma}. 
Because the small temperature dependence of $D(E_F)$ within the FM phase\cite{Park,Tanaka} 
caused by improvement of  spin allignment in the double exchange regime is almost saturated
far below $T_C$, the area of the well-screened feature is interpreted to be
proportional to the volume of the FM state.

The difference in the temperature dependence for $y$ = 0.40 and 0.25 is
understood by inspecting the phase diagram (inset of Fig. 2(b)).
With decreasing temperature, the $y$ = 0.25 film crosses a first-order
phase boundary between the COOI and FM. A small fraction of FM
appears at $\sim$ 125K followed by a gradual growth, which is
completely masked in the macroscopic resistance measurements 
by the resistivity increase of the
insulating part as the temperature is lowered. When the FM fraction reaches a
threshold for establishing a continuous path, the resistance drops abruptly.
On warming, the reverse process occurs for the insulating phase in
the metallic background. If this were to be a simple percolation
problem, the threshold should be around 0.6 for the metal phase
fraction $x_M$ \cite{last}. However, $x_M$ is estimated to be about 0.6 
even at 20K, assuming that $x_M$
at the same temperature is unity for 
$y$ = 0.40. Therefore, a simple percolation theory is not applicable
here. However, it should be noted that the threshold $x_M$ is about
0.3 - 0.4 both for cooling and warming runs. This may hint a presence of
some texture in the FM-COOI mixed state not affected by the thermal
cycling.

\begin{figure}
\includegraphics[width=8cm,clip]{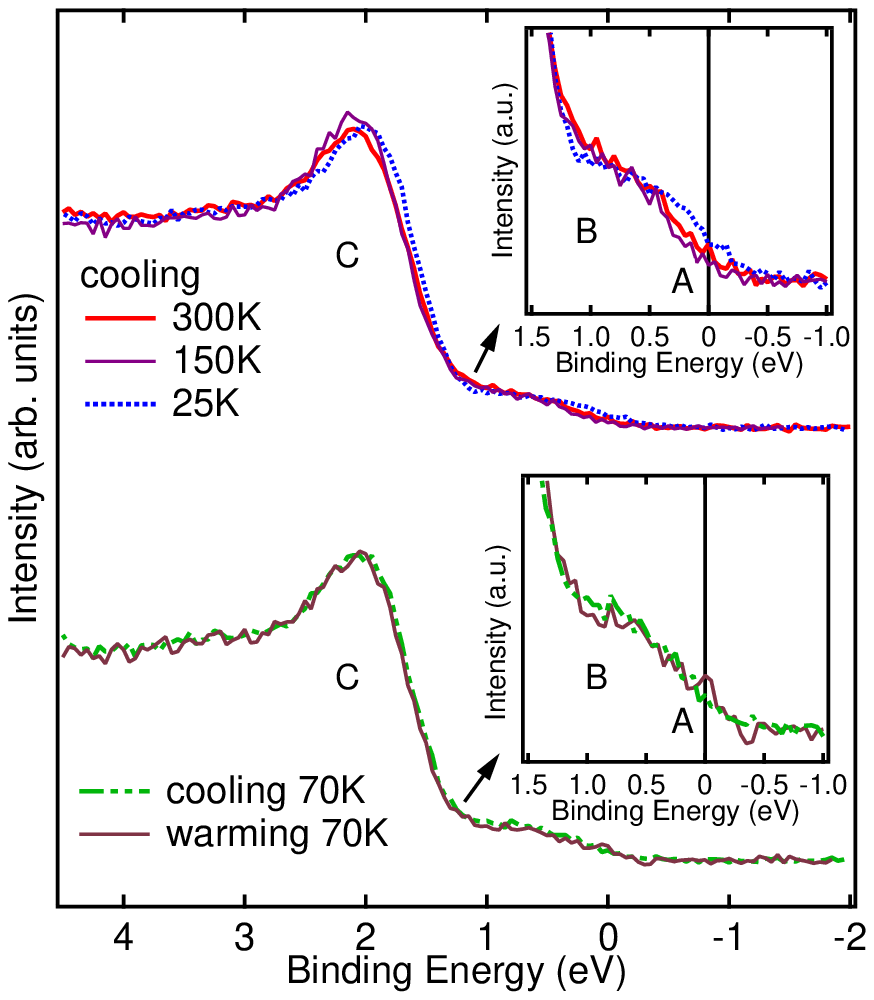}%
\caption{(Color online). Valence-band XPS data of
PCSMO ($y$ = 0.25) thin film at various temperatures.
}
\label{y25valence}
\end{figure}

These observations are also consistent with the behavior of the
valence-band spectra (Fig.\ref{y25valence}).
Comparing the metallic state (25 K) to the insulating state (300 K, 150 K),
the weight transfer from structure B to structure A is observed crossing the IMT.
The spectra at 150 K have a larger band gap ($\sim$ 0.3 eV) than those at 300 K,
reflecting the COOI nature at 150 K.
Moreover the spectra at 70 K in cooling process have smaller $D(E_F)$ compared to the spectra in the warming process.

\begin{figure}
\includegraphics[width=8cm,clip]{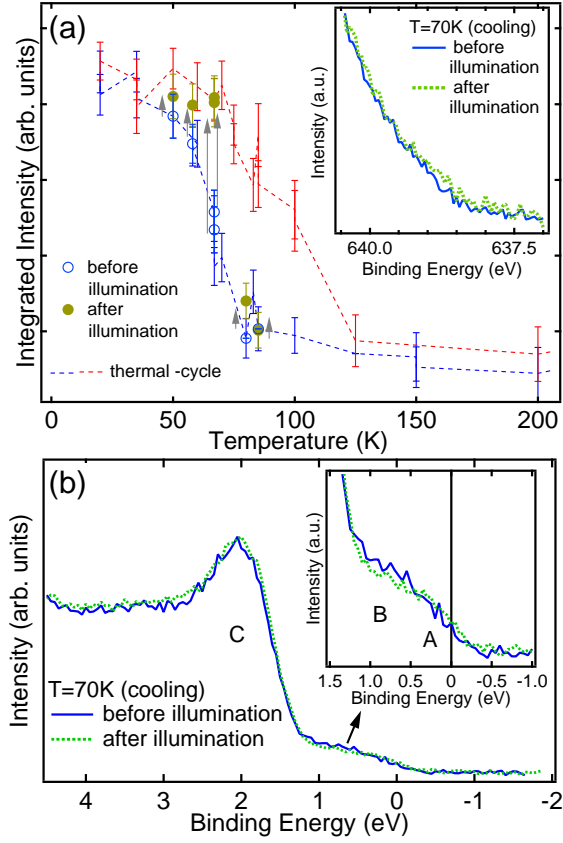}%
\caption{(Color online). (a) Integrated intensity from 638.0 eV to 640.0 eV of the XPS data for $y$ = 0.25
before and after laser illumination superposed on the temperature dependence. The arrows denote the changes caused by 
the laser illumination. The inset shows the spectra for $y$ = 0.25 at 70 K before and after the laser illumination.
(b) Valence-band XPS data for $y$ = 0.25 at 70 K before and after laser illumination.}
\label{illumination}
\end{figure}

To study the photoinduced change of the film,
the XPS data were taken for $y$ = 0.25 before and after visible light
irradiation at 50 K, 60 K, 70 K, 80K, and 85 K. 
More than 1000 laser pulses were shone on the sample.
The photoinduced effect in the integrated intensity is
superposed on the thermal hysteresis in Fig. \ref{illumination} (a).
The arrows denote the changes caused by the laser irradiation. Two features are
worth noting; (1) the laser-irradiation induces clear IMT only below a 
critical temperature (around 75K) and (2) the resulting metallic state
is identical to that reached in the warming run at the respective
temperatures [see also the insets of Fig. \ref{illumination}]. 
This transition is persistent as long as the temperature is held
constant and can
be reversed by heating above 120K 
as previously reported \cite{NTakubo}. 
In contrast, the laser illumination effect for $y$ = 0.40 has  not been observed
in all temperature range, including the insulating state.

It is clear that the PIPT in PCSMO is due to the crossing of a
first-order phase transition line below a bicritical point (inset,
Fig. 2(b)). The true phase transition point is near 75K but the system
apparently can be supercooled as much as 50K. The
photoirradiation induces charge-orbital disordered state by
charge transfer excitation, tipping the balance momnentarily to
FM. The system subsequently falls back to COOI above 75K but it is
transformed to FM below. With sufficient laser pulses, the metal volume fraction 
thus formed is as much as allowed by the equilibrium condition 
at the temperature. The PIPT is thus acting on the volume of COOI and the 
presense of the metal islands close to the percolation limit 
is not a prerequisite for the drastic resistance drop under the photoexcitation.
Even below this critical temperature, however, a
considerable amount of photon density is necessary for the
transformation to take place \cite{NTakubo}. Together with the extreme
stability of the superheated or supercooled states, it is obvious that
the COOI and FM are separated by a large energy barrier associated
with a long range order not easily destroyed by a single particle
excitation by photons.

An obvious candedate for such an order is elastic deformation. The
coupling to the Jahn-Teller type lattice distortion stabilizes the
COOI considerably while the FM is deformation free. Because the distortion is
collective and the elastic 
interaction is long range, the difficulty in overcoming the barrier is
clear, resulting in the large hysteresis. The analogy to the
martensitic transformation has been pointed out \cite{podzorov}.
Indeed, x-ray diffraction
shows that there is no sturctural change at the PI to FM transition for the 
$y$ = 0.40 film while the PI to COOI transformation for $y$ = 0.25
film is accompanied by
a lattice deformation \cite{wakabayashi}. It is to be noted
that the $y$ = 0.25 film is not 100\% FM even at the lowest
temperature. The robustness of the COOI phase in $y$ = 0.25 sample is
peculear to this composition but not due to the film being
inherently nonuniform. This
is clear by the sharp transition in the $y$ = 0.40 film.

In conclusion, we have studied the electronic structure of  PCSMO 
($y$ = 0.25, 0.40) thin fillms
by means of XPS.
In the case of $y$ = 0.40, the IMT is clearly observed on the Mn $2p$ core-level and valence-band spectra
across the PI to FM transition. 
For $y$ = 0.25, hysteresis is observed between 50 K and 125 K associated with
the COOI-FM first-order phase transition.
It is phase separated below 125 K and does not reach 100\% FM even at the lowest
temperature.
Persistent PIPT from COOI to FM is also observed for $y$ = 0.25
in the temperature range between 50 K and 70 K. After sufficient illumination,
the system is as much metallic as allowed in the equilibrium, showing
that the photons transform the bulk of COOI rather than connect the marginally
disconnected metallic islands as is proposed for CMR1.

The authors would like to acknowledge fruitful discussions with G. A. Sawatzky.
This work was supported by Grant-In-Aid
for Scientific Research (17105002, 16684010, 16204024, 16076207, 15104006) from the Japan Society for the Promotion of Science.


\begin{references}

\bibitem{Decurtins}
S. Decurtins {\it et al.},
Chem. Phys. Lett. {\bf 105}, 1 (1984).

\bibitem{Sato}
O. Sato {\it et al.},
Science {\bf 272}, 704 (1996).

\bibitem{Liu}
X. J. Liu  {\it et al.},
Phys. Rev. B {\bf 61}, 20 (2000).

\bibitem{Lindenberg}
A. M. Lindenberg {\it et al.},
Phys. Rev. Lett. {\bf 84}, 111 (2000).


\bibitem{Sokolowski}
K. Sokolowski-Tinten {\it et al.},  
Nature {\bf 422}, 287 (2003).


\bibitem{Chollet}
M. Chollet {\it et al.},
Science {\bf 307}, 86 (2005).

\bibitem{Cavalleri}
A. Cavalleri {\it et al.},
Phys. Rev. Lett. {\bf 87}, 237401 (2001).


\bibitem{Miyano}
K. Miyano {\it et al.},
Phys. Rev. Lett. {\bf 78}, 4257 (1997) and M. Fiebig {\it et al.}, Science {\bf 280}, 1925 (1998).


\bibitem{aliaga}
H. Aliaga {\it et al.},
Phys. Rev. B {\bf 68}, 104405 (2003).

\bibitem{NTakubo}
N. Takubo {\it et al.},
Phys. Rev. Lett. {\bf 95}, 017404 (2005).

\bibitem{okimoto}
Y. Okimoto  {\it et al.},
Appl. Phys. Lett. {\bf 80}, 1031 (2002).

\bibitem{endoh}
Y. Endoh  {\it et al.},
Phys. Rev. Lett. {\bf 82}, 4328 (1999).


\bibitem{Tomioka}
Y. Tomioka and Y. Tokura,
Phys. Rev. B {\bf 66}, 104416 (2002).



\bibitem{Horiba}
K. Horiba {\it et al.},
Phys. Rev. Lett {\bf 93}, 236401 (2004).

\bibitem{Tanaka}
H. Tanaka  {\it et al.},
Phys. Rev. B {\bf 73}, 094403 (2006).



\bibitem{Chainani}
A. Chainani  {\it et al.},
Phys. Rev. B {\bf 56}, R15513 (1997).

\bibitem{Sekiyama}
A. Sekiyama  {\it et al.},
Phys. Rev. B {\bf 59}, 15528 (1999).


\bibitem{Sarma}
D. D. Sarma {\it et al.}, 
Phys. Rev. Lett {\bf 93}, 097202 (2004).


\bibitem{Park}
J.-H. Park  {\it et al.},
Phys. Rev. Lett. {\bf 76}, 4215 (1996). 


\bibitem{last}
B. J. Last and D. J. Thouless,
Phys. Rev. Lett. {\bf 27}, 1719 (1071).

\bibitem{podzorov}
V. Podzorov {\it et al.},
Phys. Rev. B{\bf 64}, 140406(R) (2001).

\bibitem{wakabayashi}
Y. Wakabayashi - private communication.

\end{references}
\end{document}